\documentclass[aps,prl,superscriptaddress,floatfix,showpacs]{revtex4}

\usepackage{graphicx}
\usepackage{epsfig}
\usepackage{amssymb}
\usepackage{amsmath}

\begin{document}

\title{NMR Time Reversal Experiments in Highly Polarised Liquid $^3$He-$^4$He Mixtures}

\author{E.~Baudin}
\affiliation{Laboratoire Kastler Brossel, Ecole Normale
Sup{\'e}rieure; CNRS; UPMC; 24 rue Lhomond, F75005 Paris, France}
\author{M.~E.~Hayden}
\affiliation{Physics Department, Simon Fraser University, 8888
University Drive, Burnaby BC, Canada V5A 1S6}
\author{G. Tastevin}
\author{P.~J.~Nacher}
\affiliation{Laboratoire Kastler Brossel, Ecole Normale
Sup{\'e}rieure; CNRS; UPMC; 24 rue Lhomond, F75005 Paris, France}

\begin{abstract}
Long-range magnetic interactions in highly magnetised liquids (laser-polarised $^3$He-$^4$He dilute mixtures at 1 K in our experiment) introduce a significant non-linear and non-local contribution to the evolution of nuclear magnetisation that leads to instabilities during free precession. 
We recently demonstrated that a multi-echo NMR sequence, based on the magic sandwich pulse scheme developed for solid-state NMR, can be used to stabilise the magnetisation against the effect of distant dipolar fields. Here, we report investigations of echo attenuation in an applied field gradient that show the potential of this NMR sequence for spin diffusion measurements at high magnetisation densities. 
\end{abstract}
\pacs{76.60.Jx, 76.60.Lz, 82.56-b, 82.56.Jn, 67.65+z, 75.40-Gb\\
Accepted for publication in {\it J. Low Temp. Phys.}}

\maketitle

\section{Introduction}
Long-range dipolar interactions between the local magnetisation and that of
the remainder of a liquid sample play a significant role in determining nuclear spin dynamics at high magnetisation
densities. They introduce a non-local contribution to the evolution of magnetisation and give rise to complex non-linear behaviour. Spectacularly
abrupt decays of NMR free induction signals have been observed at large tip
angles in several laser-polarised liquid systems \cite{nacher00}, and have been explained by the
development of unstable inhomogeneous magnetisation patterns that grow
exponentially in time \cite{jeener99}. The onset of precession instabilities is a generic
feature of highly magnetised systems, and is related to turbulent spin motion \cite{jeener99}. 
It is potentially relevant to understanding the dynamics of polarised
quantum fluids, and is becoming increasingly important in high-field $^{1}$H-based 
NMR spectrometry where chaotic spatiotemporal behaviour occurs under
the joint action of distant dipolar fields and strong radiation damping \cite{lin00}.

We recently demonstrated that the magic sandwich (MS), a sophisticated
pulse sequence designed to refocus the magnetisation lost through broadening by local dipolar fields in solids, can be 
successfully used to fight the deleterious effect of distant dipolar fields
in highly magnetised liquids and to investigate the spatiotemporal
development of instabilities \cite{hayden07}. We also demonstrated that 
it is possible to achieve dynamic stabilisation of transverse
precession in laser-polarised He mixtures using a series
of MS cycles to periodically refocus the dipolar interactions. In this
paper, we report preliminary results that establish the potential of
multiple-echo MS-based NMR sequences for spin diffusion measurements
at high magnetisation densities, and confirm the fact that conventional 
spin echo techniques completely fail to achieve this goal \cite{piegay00}.

\section{Experimental setup}

Polarised samples are prepared by condensation of optically pumped $^{3}$He
gas into superfluid $^{4}$He in a 1 metre-long Pyrex glass cell. $^{3}$He gas is
continuously injected (at mbar pressures with a flow rate of order 0.1~$\mu$mol/s)        
into a room temperature volume where laser optical
pumping is performed, as described elsewhere \cite{piegay02}. It then flows down a narrow tube into 
a 0.44~cm$^{3}$ spheroidal volume partly filled 
with liquid $^{4}$He at temperature $T\simeq$ 1.15~K. Cs coatings are used to avoid
wall relaxation, and bulk dipole-dipole relaxation times
reach several hours in our dilute samples ($^{3}$He molar fractions $X=$~1-5\%) \cite{piegay02}. 
By adjusting the $^{3}$He molar fraction and nuclear
polarisation (that can be as high as 40\%), we control the magnetisation
density $M$ in the sample.
In the following, $M$ is conveniently
reported in terms of a dipolar frequency $F_{\textrm{dip}}=\gamma B_{\textrm{dip}}/2\pi $ 
associated with the characteristic dipolar field $B_{\textrm{dip}}=\mu _{0}M$, where $\mu _{0}$ 
is the permeability of free space and $\gamma $ the $^{3}$He
gyromagnetic ratio ($\gamma/2\pi=32.4$~MHz/T). 
NMR excitation by 90$^{\circ}$ rf pulses induces 
an irreversible loss of this out-of-equilibrium
magnetisation, so that a new batch of polarised gas has to be condensed for
each experiment.

The cryogenic apparatus includes a commercial aluminum and fiberglass dewar and a homemade non-magnetic insert. The lower end of the cell
lies inside a stainless-steel vacuum can, and is thermally anchored to a
copper 1~K-pot via bundles of enameled copper wires that do not allow for eddy current
loops.

Low frequency NMR (74~kHz) is performed in a 2.3~mT magnetic field (shimmed
to 20~ppm over the sample) using crossed coils. 
The receive coil is capacitively tuned with a very low $Q$ factor ($\simeq 2$)
to avoid radiation damping. The transmit coil is actively
shielded to avoid Joule heating by eddy currents induced in the surrounding metallic parts during intense cw rf excitation \cite{bidinosti05}. With the shield coil in place, the rf field outside of the
transmit coil is dramatically reduced and, e.g., the measured heat load on the 1~K-pot is more than two orders of magnitude lower. Shielding also minimises field distortions and improves rf homogeneity ($\pm 200$~ppm over the sample). In-house PC-controlled analog hardware allows for flexible control of hard pulse sequences. 

\section{Echo trains and diffusion-induced attenuation}

Spin diffusion coefficients are usually measured by monitoring the
attenuation of echo amplitudes in the presence of a uniform applied
gradient. Conventional spin echo techniques are inefficient in highly magnetised liquids. The (quadratic) effect of dipolar
interactions actually remains unchanged throughout the sample when a 180$^{\circ}$ rf pulse is applied to rotate the local magnetisation. 
The decay of the average transverse magnetisation is thus expected to remain dominated by dipolar-induced 
instabilities at high magnetisation densities. Indeed, early attempts to generate single Hahn spin echoes and Carr-Purcell spin echo trains in highly
polarised $^{3}$He-$^{4}$He dilute mixtures have led to the observation of multiple
echoes \cite{tastevin03} and of non-exponential decay \cite{piegay00}, respectively.

\begin{figure}[ht]
\begin{center}
\includegraphics[width=13cm]{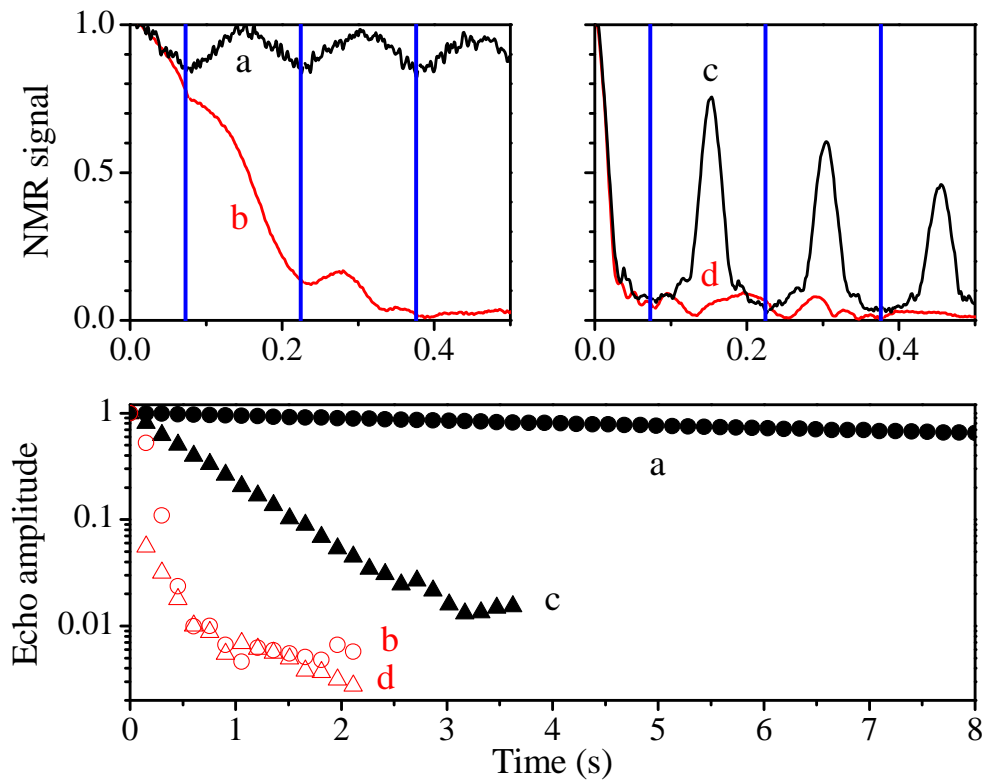}
\caption{\label{Fig1} Echo trains obtained using a CPMG sequence with $2\tau$=151~ms in mixtures with $X$=2.4-2.9\% at $T$=1.16~K, for various gradients and magnetisations. 
Upper plots: FID and first 3 echoes for $G$=0.47~$\rm{\mu}$T/cm (left) and $G$=3.34~$\rm{\mu}$T/cm (right), at low magnetisations (a: $F_{\textrm{dip}}$=0.4~Hz, c: $F_{\textrm{dip}}$=0.5~Hz) or high magnetisations (b: $F_{\textrm{dip}}$=11.5~Hz, d: $F_{\textrm{dip}}$=25~Hz). 180$^{\circ}$ rf pulses are indicated by vertical lines. 
Lower plot: Semi-log plot of echo amplitudes versus time, for these echo trains (filled/open symbols: low/high $F_{\textrm{dip}}$; circles/triangles: weak/strong $G$). 
For low $F_{\textrm{dip}}$, decay rates $\Gamma$=0.054 and 1.42~s$^{-1}$ are extracted from echo trains a and c, respectively. 
}
\end{center}
\end{figure}

Figure \ref{Fig1} illustrates the differences in time evolution of signals obtained with a conventional spin echo sequence at high and low magnetisation densities for two different values of the applied gradient $G$. 
The Carr-Purcell-Meiboom-Gill (CPMG) spin echo sequence $90_y^{\circ }-\tau-(180_x^{\circ }-2\tau )_{n}$ 
is used to reduce cumulated phase errors \cite{meiboom58}. 
At low magnetisations (traces a and c), spin echoes have the shapes 
that are expected given the sample geometry and the gradient strengths. Mono-exponential decays of the echo amplitudes are observed. 
At high magnetisations (traces b and d), the echo shapes are strongly altered and the much faster decay is no longer exponential, as previously reported \cite{piegay00,nacher02}. 
Since 180$^{\circ}$ pulses only remove the linear phase dispersion introduced by the applied
gradient, dipolar instabilities quickly take over and induce a very strong signal decay. 
At low $G$, minor refocussing is achieved by the 180$^{\circ}$ pulses and the echo envelope is reminiscent of a typical dipolar-induced signal decay. 
At high $G$, the large gradient-induced frequency spread over the sample ($\sim 10 \times F_{\textrm{dip}}$) dominates the very short measured FID lifetime. 
However, irreversible signal loss occurs during subsequent free evolution due to the development of unstable magnetisation patterns.  

Dipolar-induced phase dispersion can be refocussed by continuous application of
appropriately phased intense rf pulses that mix the longitudinal and transverse
components of the magnetisation in such a way that it undergoes a
time-reversed evolution, at half the pace of free evolution \cite{rhim71}. 
Successful implementation and use of the original magic sandwich pulse scheme in the context of liquid NMR has been recently reported \cite{hayden07}. 
We use a slightly modified MS cycle, $90^{\circ}_{y}(360^{\circ}_{x},$-$360^{\circ}_{x})_{n}90^{\circ}_{y}$, that incorporates an additional 180$^{\circ}$ rotation to refocus the gradient-induced linear phase dispersion. 
Periodic refocussing is achieved using a \lq repeated magic sandwich\rq \ (RMS)
sequence $90^{\circ}-\tau -(\rm{MS}-2\tau )_{n}$, where the duration of the MS cycle is $4\tau$ \cite{hayden07,baudin07}. 
The RMS sequence is thus analogous to a conventional spin echo sequence
where the 180$^{\circ}$ pulses have been replaced by MS cycles. It leads to the formation of
so-called magic echoes with a time period equal to $6\tau$. 

\begin{figure}[ht]
\begin{center}
\includegraphics[width=13cm]{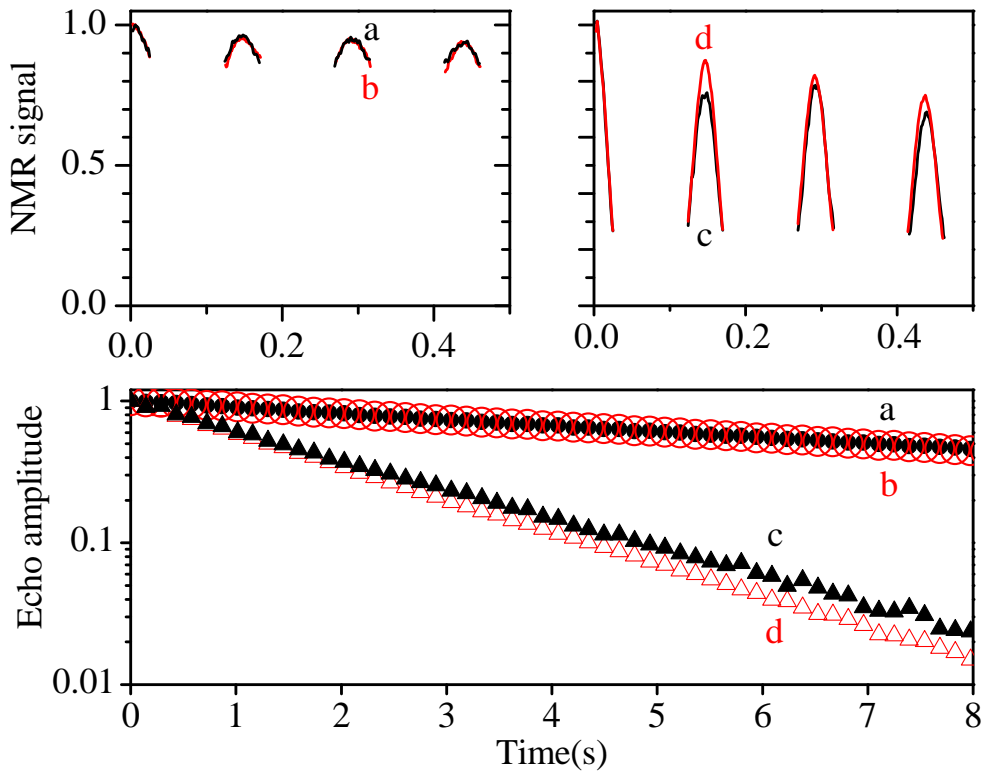} 
\caption{\label{Fig2} Echo trains obtained using a RMS sequence with $6\tau$=145~ms in mixtures with $X$=2.4-3.0\% at $T$=1.16~K, for various gradients and magnetisations. 
Upper plots: FID and first 3 echoes for $G$=0.89~$\rm{\mu}$T/cm (left) and $G$=3.34~$\rm{\mu}$T/cm (right), at low magnetisations (a: $F_{\textrm{dip}}$=0.8~Hz, c: $F_{\textrm{dip}}$=0.9~Hz) and at high magnetisations (b: $F_{\textrm{dip}}$=10~Hz, d: $F_{\textrm{dip}}$=13~Hz). Note that the low and high $F_{\textrm{dip}}$ traces in the upper-left-hand plot are nearly identical. No signal is acquired during the MS cycles. 
Lower plot: Semi-log plot of the echo amplitudes for these echo trains (filled/open symbols: low/high $F_{\textrm{dip}}$; circles/triangles: weak/strong $G$). 
For low $F_{\textrm{dip}}$, decay rates $\Gamma$=0.099 and 0.49~s$^{-1}$ for a and c, respectively. For high $F_{\textrm{dip}}$, $\Gamma$=0.098 and 0.52~s$^{-1}$ for b and d, respectively.
}
\end{center}
\end{figure}

Figure \ref{Fig2} shows magic echo trains obtained under conditions similar to those of Fig. \ref{Fig1}. 
All echoes have the expected shapes and exhibit monoexponential decay at a rate that does not depend on $F_{\textrm{dip}}$. 
NMR signals are efficiently recovered at all magnetisation densities since frequent time reversal heals spatial variations of magnetisation that develop during periods of free evolution. 

For both CPMG and RMS echo trains, experimental decay rates $\Gamma$ are obtained from an exponential fit to the decay of the squared amplitude of the Fourier transform of the band-pass filtered NMR signals. This method preserves the accuracy of the analysis down to low signal-to-noise ratios. 

\section{Discussion}
Our results confirm that
CPMG sequences cannot be used to probe spin diffusion in highly magnetised liquids, due to the dramatic dipolar-induced signal losses (Fig.~\ref{Fig1}, traces b and d). 
In the stable linear regime (at low $F_{\textrm{dip}}$), the diffusion-induced decay rate for spin echo amplitudes is \cite{stejskal65}:
\begin{equation}
\label{eq1}
\Gamma ^{\textrm{CPMG}}=D(2\tau \gamma G)^{2}/12,
\end{equation}
where $D$ is the spin diffusion coefficient in the mixture.
The quadratic dependence of $\Gamma ^{\textrm{CPMG}}$ on $\tau$ and $G$ has been checked in a previous series of experiments, in which it was demonstrated that $D$ can be accurately measured with this method in our apparatus \cite{piegay00}. 
Here, the two decay rates $\Gamma$ extracted from traces a and c in Fig.~\ref{Fig1} yield $D=1.54 \times 10^{-3}$~cm$^2$/s at $X=2.8$\%, in good agreement with published values \cite{opfer68,piegay00}. 

The RMS sequence appears to be a lot more robust than the CPMG sequence against dipolar effects. Over the range of experimental conditions we have explored, the shape and attenuation of magic echoes are insensitive to $F_{\textrm{dip}}$. The expected diffusion-induced decay rate for spin echo amplitudes is \cite{baudin07}: 
\begin{equation}
\label{eq2}
\Gamma ^{\textrm{RMS}}=\frac{7}{27} D(6\tau \gamma G)^2/12,
\end{equation}
which is less than $\Gamma ^{\textrm{CPMG}}$ because the dephasing action of $G$ is arrested during each MS cycle \cite{baudin07}. Assuming that the decay rates measured for the strongest $G$ only result from diffusion-induced attenuation (i.e., $\Gamma=\Gamma ^{\textrm{RMS}}$), traces c and d in Fig.~\ref{Fig2} combined with Eq.~\ref{eq2} yield $D=2.4  \times 10^{-3}$~cm$^2$/s at $X=2.5$\%, which is 40\% higher than expected. Decay rates at zero applied gradient (obtained from direct measurements at $G$=0 or from extrapolation of RMS data at various $G$) are also higher than expected from CPMG measurements (0.13-0.24~s$^{-1}$ instead of 0.033~s$^{-1}$). The physical origin of this additional damping has not yet been identified. We believe that rf heating of the sample (of order 1~mK/s during RMS sequences) has negligible influence on signal decay. In our experiment, evaporation hardly modifies the $^3$He molar fraction (a 140~ppm/s relative decrease of $X$ is estimated from thermodynamic data \cite{nacher94}).  

Additional signal losses incurred during RMS sequences, that are observed to be
$F_{\textrm{dip}}$-independent, may be caused by minor rf amplitude and timing imperfections that will soon be addressed.  
The present work clearly demonstrates that robust RMS sequences have the potential to
enable one to perform accurate diffusion measurements in highly magnetised liquids. More
generally, the efficient MS sequence provides a valuable time-reversal tool
to investigate the complex non-linear NMR dynamics that can be encountered in
polarised quantum fluids such as Bose-Einstein condensates, superfluid $^3$He, degenerate $^3$He-$^4$He mixtures, and 2D hydrogen gases.

\end{document}